\documentclass[twocolumn]{aastex631}

\usepackage{amsmath}
\usepackage{comment}

\shorttitle{Skewed Microlensing in Dragon Arc}
\shortauthors{Broadhurst et al.}

\graphicspath{{./}{figures/}}

\begin{document}

\title{Dark Matter distinguished by skewed microlensing in the ``Dragon Arc".}

\author[0000-0002-8785-8979]{Tom Broadhurst}
\email{tom.j.broadhurst@gmail.com}
\affiliation{Donostia International Physics Center, DIPC, Basque Country, San Sebasti\'an, 20018, Spain}
\affiliation{Department of Physics, University of Basque Country UPV/EHU, Bilbao, Spain}
\affiliation{Ikerbasque, Basque Foundation for Science, Bilbao, Spain}

\author[0000-0002-4490-7304]{Sung Kei Li}
\affiliation{Department of Physics, The University of Hong Kong, Pokfulam Road, Hong Kong}

\author[0000-0003-1276-1248]{Amruth Alfred}
\affiliation{Department of Physics, The University of Hong Kong, Pokfulam Road, Hong Kong}

\author[0000-0001-9065-3926]{Jose M. Diego}
\affiliation{IFCA, Instituto de F\'isica de Cantabria (UC-CSIC), Av. de Los Castros s/n, 39005 Santander, Spain}

\author[0009-0008-5791-1012]{Paloma Morilla}
\affiliation{Department of Physics, University of Basque Country UPV/EHU, Bilbao, Spain} 
\affiliation{Donostia International Physics Center, DIPC, Basque Country, San Sebasti\'an, 20018, Spain}

\author[0000-0003-3142-997X]{Patrick L. Kelly}
\affiliation{Minnesota Institute for Astrophysics, University of Minnesota, 116 Church St. SE, Minneapolis, MN 55455, USA}

\author[0000-0002-4622-6617]{Fengwu Sun} 
\affiliation{Center for Astrophysics $|$ Harvard \& Smithsonian, 60 Garden St., Cambridge, MA 02138, USA}
\affiliation{Steward Observatory, University of Arizona, 933 N. Cherry Ave., Tucson, AZ 85721, USA}

\author[0000-0003-3484-399X]{Masamune Oguri} 
\affiliation{Center for Frontier Science, Chiba University, 1-33 Yayoi-cho, Inage-ku, Chiba 263-8522, Japan}
\affiliation{Department of Physics, Graduate School of Science, Chiba University, Chiba 263-8522, Japan} 

\author[0000-0002-1681-0767]{Hayley Williams}
\affiliation{Minnesota Institute for Astrophysics, University of Minnesota, 116 Church St. SE, Minneapolis, MN 55455, USA}

\author[0000-0001-8156-6281]{Rogier Windhorst}
\affiliation{School of Earth and Space Exploration, Arizona State University, Tempe, AZ 85287-1404, USA}

\author[0000-0002-0350-4488]{Adi Zitrin}
\affiliation{Department of Physics, Ben-Gurion University of the Negev, PO Box 653, Be’er-Sheva 8410501, Israel}

\author[0000-0001-5474-4716]{Katsuya T. Abe}
\affiliation{Center for Frontier Science, Chiba University, 1-33 Yayoi-cho, Inage-ku, Chiba 263-8522, Japan}

\author[0000-0003-1060-0723]{Wenlei Chen}
\affiliation{Department of Physics, Oklahoma State University, 145 Physical Sciences Bldg, Stillwater, OK 74078, USA}

\author[0000-0001-7440-8832]{Yoshinobu Fudamoto} 
\affiliation{Center for Frontier Science, Chiba University, 1-33 Yayoi-cho, Inage-ku, Chiba 263-8522, Japan}

\author[0009-0006-6911-2299]{Hiroki Kawai}
\affiliation{Department of Physics, School of Science, The University of Tokyo, Bunkyo, Tokyo 113-0033, Japan}
\affiliation{Center for Frontier Science, Chiba University, 1-33 Yayoi-cho, Inage-ku, Chiba 263-8522, Japan}

\author[0000-0003-4220-2404]{Jeremy Lim}
\affiliation{Department of Physics, The University of Hong Kong, Pokfulam Road, Hong Kong}

\author[0000-0002-5248-5076]{Tao Liu}
\affiliation{Department of Physics, The Hong Kong University of Science and Technology, Hong Kong S.A.R., P.R.China}
\affiliation{Jockey Club Institute for Advanced Study, The Hong Kong University of Science and Technology, Hong Kong S.A.R., P.R.China}

\author[0000-0002-7876-4321]{Ashish K. Meena}
\affiliation{Department of Physics, Ben-Gurion University of the Negev, PO Box 653, Be’er-Sheva 8410501, Israel}

\author[0000-0003-0942-817X]{Jose M. Palencia}
\affiliation{IFCA, Instituto de F\'isica de Cantabria (UC-CSIC), Av. de Los Castros s/n, 39005 Santander, Spain}

\author[0000-0001-7575-0816]{George F. Smoot}
\affiliation{Donostia International Physics Center, DIPC, Basque Country, San Sebasti\'an, 20018, Spain}
\affiliation{Department of Physics and Institute for Advanced Study, The Hong Kong University of Science and Technology, Hong Kong}
\affiliation{Paris Centre for Cosmological Physics, APC, AstroParticule et Cosmologie, Universit´e de Paris, CNRS/IN2P3, CEA/lrfu, 10, \\ rue Alice Domon et Leonie Duquet, 75205 Paris CEDEX 13, France emeritus}
\affiliation{Physics Department, University of California at Berkeley, CA 94720, Emeritus}

\author[0000-0002-6039-8706]{Liliya L.R. Williams}
\affiliation{Minnesota Institute for Astrophysics, University of Minnesota, 116 Church St. SE, Minneapolis, MN 55455, USA}











\begin{abstract}

 Microlensed stars recently discovered by JWST \& HST follow closely the winding critical curve of A370 along all sections of the ``Dragon Arc"  traversed by the critical curve. These transients are fainter than $m_{AB}>26.5$, corresponding to the Asymptotic Giant Branch (AGB) and microlensed by diffuse cluster stars observed with $\simeq 18M_\odot/pc^2$, or about $\simeq 1$\% of the projected dark matter density. Most microlensed stars appear along the inner edge of the critical curve, following an asymmetric band of width $\simeq 4$kpc that is skewed by $-0.7\pm0.2$kpc. Some skewness is expected as the most magnified images should form along the inner edge of the critical curve with negative parity, but the predicted shift is small $\simeq -0.04$kpc and the band of predicted detections is narrow, $\simeq 1.4$kpc. Adding CDM-like dark halos of $10^{6-8}M_\odot$ broadens the band as desired but favours detections along the outer edge of the critical curve, in the wrong direction,  where sub-halos generate local Einstein rings. Instead, the interference inherent to ``Wave Dark Matter" as a Bose-Einstein condensate ($\psi$DM) forms a symmetric band of critical curves that favours negative parity detections. A de Broglie wavelength of $\simeq 10$pc matches well the observed $4$kpc band of microlenses and predicts negative skewness $\simeq -0.6$kpc, similar to the data. The implied corresponding boson mass is $\simeq 10^{-22}$eV, in good agreement with estimates from dwarf galaxy cores when scaled by momentum. Further JWST imaging may reveal the pattern of critical curves by simply ``joining the dots" between microlensed stars, allowing wave corrugations of $\psi$DM to be distinguished from CDM sub-halos. 

\end{abstract}

\keywords{Galaxies: Clusters: individual (Abell 370) -- Gravitational Lensing: Strong, Micro}
\section{Introduction} \label{sec:intro}

 Detection of microlensed stars at cosmological distances has become routine in cadenced JWST or deep Hubble imaging \citep{Kelly_2023a}. These are typically found where lensed galaxies are bisected by the radial or tangential critical curves of massive lensing clusters. The host galaxies showing microlensing are preferentially of modest redshift, which is to be expected given the rapidly increasing luminosity distance with redshift, so that in practice super giant stars become much harder to detect at $z>2$. These microlensed stars are recognised by flux variation, including clear cases of caustic crossing where the magnification saturates for some hours as the caustic crosses the finite area stellar disk and then suddenly disappears on the other side of the caustic. The first example at cosmological distances, ``Icarus", was discovered serendipitously \citep{Kelly2018} with other cases subsequently discovered \citep{Rodney2018,Chen2019}. It has become clear from these light curves that the relatively modest projected mass density of diffuse cluster stars observed as intra-cluster light (ICL) is sufficient to account for the light curve variation, with a projected stellar surface mass density of typically $10M_\odot/pc^2$, representing only 1\% of the total projected column of matter at the location of stars on the Einstein radius, where dark matter dominates. This consistency with the observed surface brightness of the ICL severely restricts any additional non-stellar microlensing mass, for example LIGO-like black holes, for which frequent, low amplitude brightness fluctuations should be seen, unlike the data \citep{Diego_2018}.
  
Related to these transients we have also recognised a puzzling class of ``one sided" unresolved sources, noticeably offset from the cluster critical curves by up to a few kpc \citep{Meena2023,Diego_2023}, with a conspicuous absence of the predicted counter image on the other side of the cluster critical curve. This absence may be blamed on microlensing by ICL stars in the cluster or millilensing by a dark perturbing halo or dim globular cluster. One of these transients is near the center of an arc behind A370, at z=1.27, unrelated to the Dragon Arc, detected as part of the Hubble ``Flashlights" program \citep{Kelly_2023a} and noted for its puzzlingly large offset of 3 kpc ($\sim 0.5\arcsec$) from the tangential critical curve of A370 \citep{Meena_2023}. We note this transient appears ``inside" the critical curve of the cluster A370 with negative image parity, as is also the case for three other microlensed transients detected recently by JWST in fold arcs behind other lensing clusters including A2744 (z=2.65) \citep{Chen_JWST}, ``Mothra" (z=2.09)\citep{Diego_Mothra} and ``Quyllur"(z=2.19) \citep{Diego_Quyllur}, with offsets of $\simeq 0.1\arcsec$ to the negative parity side of the respective tangential critical curves.  Repeat imaging of the higher redshift ``one-sided"  cases has not found time variation, disfavouring microlensing and pointing to unresolved star clusters that may be ``millilensed" by dark perturbations \citep{Dai_2020,Diego2024_3M}. Millilensing, unlike microlensing, has a relatively low density of critical curves and so high magnification caustic crossing is unlikely and thus insufficient for detection of a distant star, but instead a can modestly magnified luminous star cluster, sufficient for detection and leave it unresolved and without time variation, as observed. Thus millilensed star clusters may provide a more plausible explanation for the higher redshift unresolved, persistent "one-sided" cases so that excessive microlensing magnification need not be invoked, such as ``Earendel"(z=6.2) \cite{Welch_2022} or ``Godzilla" \citep{Godzilla}(z=2.4) for which a dense, young star cluster explanation is preferred spectroscopically \cite{Pascale} and plausibly also for ``MACS0647-star-1" (z=4.8) \cite{Furtak}. The milli-lensing scale corresponds to sub-halos of $10^{6-8}M_\odot$ and the absence of detected starlight light even in deep JWST images, excludes normal globular clusters, preferring dark matter possibilities such as dark, low mass CDM halos or the pervasive lensing corrugations of Wave Dark Matter as a Bose-Einstein Condensate, $\psi$DM, \citep{Chan_2020, Kawai_2022, Amruth_2023}. Here we will explore the combination of dark matter substructure induced millilensing with the level microlensing by stars in the cluster comprising the diffuse intracluster light. 


\begin{figure*}[ht!]
    \centering
    \includegraphics[width = \textwidth]
    {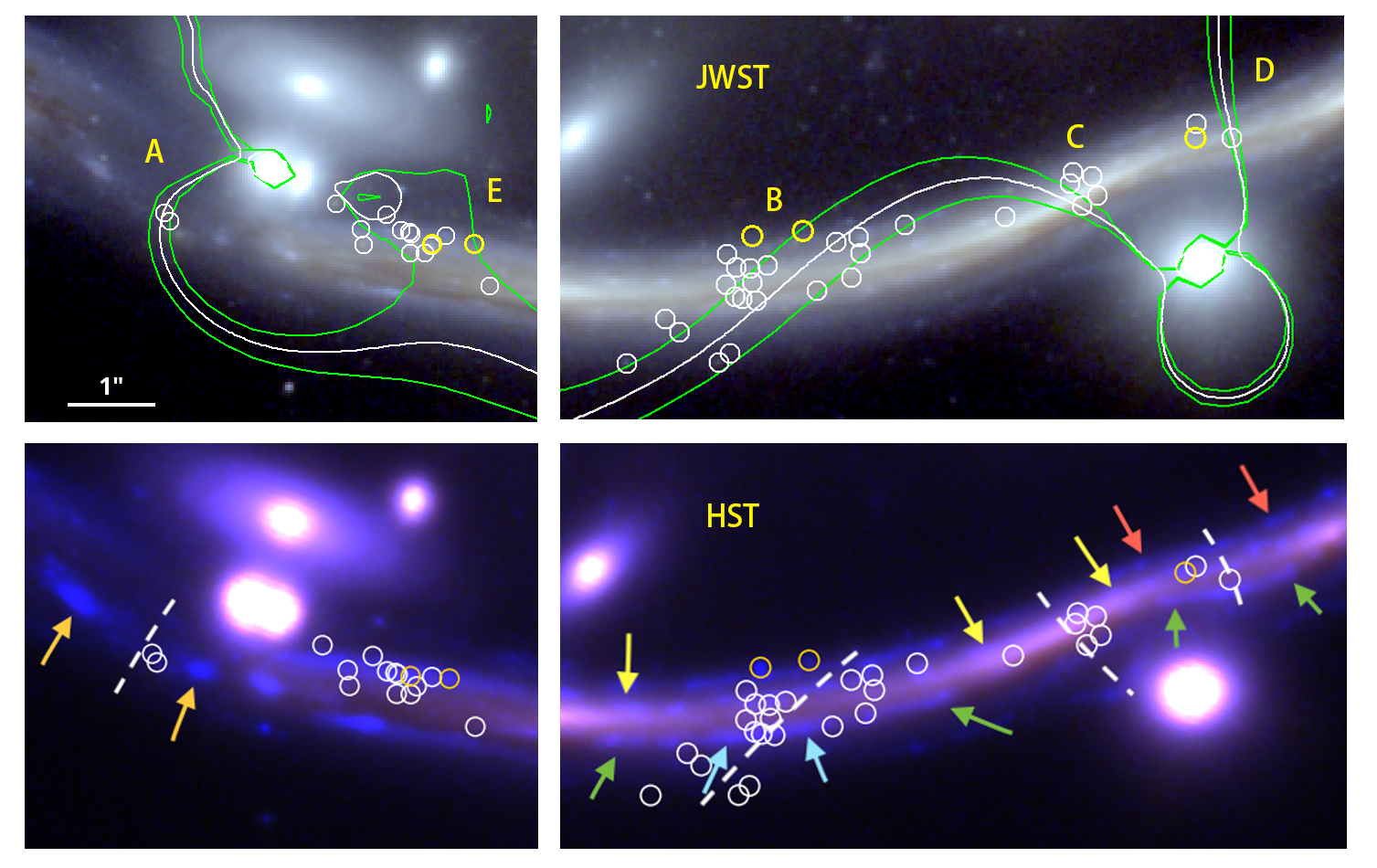}
    \caption{The JWST transients found in the Dragon arc \citep{Fudamoto_2024} shown as white circles, plus those found by our Flashlights/Hubble program \citep{Kelly_2023a} as yellow circles, revealing the microlenses follow closely the critical curve, appearing along all four intersections of the Dragon Arc with the tangential critical curve, labelled A,B,C,D, implying these events are highly magnified. The tangential critical curve in the upper panels is indicated by the white line and bracketed by green lines showing the $\mu= \pm 100$ magnification contours from the free-form strong lensing WSLAP+ code \citep{Diego2024_3M} using over 90 multiply lensed galaxies and made independently, prior to the discovery of microlensing here. The right hand panel shows the accuracy of the critical curve at the A \&B intersections, where it agrees precisely with the symmetry of internal features identified in Hubble \& JWST. The lower panels show the deep Hubble images from our Flashlights program showing many internal pairs of features, some of which are marked with arrows here and colour coded to indicate where reflection symmetry pins down the path of the winding critical curve of A370 as it crosses the Dragon Arc. This also demonstrates the good agreement of this model-independent critical path with the WSLAP+ model shown in the upper panel.}
    \label{fig: DragonArc_Transients}
\end{figure*}

  Pervasive dark matter substructure is predicted for Wave Dark Matter, $\psi$DM, by de Broglie scale wave interference for dark matter as a Bose-Einstein condensate, where the boson mass, $m_\psi$, is the only free parameter. These fluctuations are equally positive and negative as they range over constructive to destructive interference with full density modulation on the de Broglie scale, $\lambda_{\psi}$ \citep{Schive_2014}. Importantly this scale depends on the momentum of the DM, via the Uncertainty Principle, $\lambda_{\psi}=\hbar/{m_{\psi} \sigma}$, so that the largest scale of modulation occurs for lower mass galaxies where it is estimated the core radius of common dwarf spheroidals (dSph) to be about $0.5$Kpc, corresponding to a boson mass of about $m_\psi=10^{-22}eV$ \citep{Schive_2014, Veltmaat_2018,Niemeyer_2020,Pozo_2024}. Given this boson mass we can anticipate a de Broglie scale of about $\simeq 15$pc for galaxy clusters where momentum is highest, or sub-arcsecond scales for lensing. A dense band of corrugated critical curves is predicted, modulated on the de Broglie scale \citep{Chan_2020, Amruth_2023}, instead of the narrow continuous tangential and radial critical curves for smooth dark matter. This behaviour we have previously shown can account quantitatively for the commonly detected ``flux anomalies" typical of multiply lensed QSO's, showing $\simeq 30$\% discrepancies in fluxes relative to smooth lens models \citep{Nierenberg} that are consistent with the level of local flux anomaly predicted for $\psi$DM \citep{Amruth_2023} due to the pervasive density fluctuations affecting image magnification locally. Furthermore, the milli-arcsecond scale ``positional anomalies" now recognised in high resolution radio images of compact lensed sources \citep{Hartley} are also quantitatively accounted for $\psi$DM density perturbations affecting lensing deflection angles locally \citep{Amruth_2023}. Such anomalies may also imply a population of dark CDM sub-halos perturbations, if not erased over time by tidal forces, but the statistical effects of $\psi$DM are very different from CDM sub-halos as small perturbing Einstein rings should be seen along the outer edge of the critical curve for CDM-like subhalos \citep{williams2023flashlights,Abe_2024} and are examined here using the Dragon Arc.
  
  The large set of over 40 microlensed stars is now reported by \citep{Fudamoto_2024} within the original ``Dragon Arc", the first known giant arc, is a relatively low redshift, $z=0.735$, bright spiral galaxy lensed by the massive cluster Abell 370 ($z=0.37$). These transients add to the 10 blue giants recently detected by Hubble in the  Dragon Arc, as part of the Flashlights program \citep{Kelly_2023a,Li_2024}. This new discovery of abundant microlensed stars in the Dragon Arc allows tests for dark matter sub-structure in exciting detail \citep{Diego2024_3M}. In section \S1 we first model the observed number counts of events to identify the stellar class responsible for the majority of JWST events and their magnification, by combining statistical microlensing with stellar synthesis code calculations, following the pioneering JWST predictions for microlensing at cosmological distances \citet{Windhorst_2018b,Meena_2022}. In \S2 we explore the simplest case of smooth dark matter for the cluster with the observed level of microlensing implied by the diffuse intra-cluster light (ICL) measured adjacent to the Dragon arc. In \S3 we add CDM subhalos and we also explore $\psi$DM with our perturbation generating code, for comparison with the width of the microlensing band and its asymmetry. Note at the redshift of the lensing cluster A370, $z=0.37$, where 1" corresponds to a scale of 5.2Kpc in the lens plane for standard cosmology.
  
\section{Origin of microlensed stars in the Dragon Arc.}

  We first examine the brightness distribution of the microlensed events detected by JWST, as recently reported by \citet{Fudamoto_2024} found with NIRCam imaging, firstly Cycle-1 GTO-1208 (the CAnadian NIRISS Unbiased Cluster Survey, CANUCS, PI: C. Willot; and then Cycle-2 GO-3538 (PI: E. Iani) targeting Abell 370 ($z=0.375$) using multiple NIRCam filters, separated by  $\sim1\,{\rm year}$. The number counts of these detections are plotted in Figure~\ref{fig: DragonArc_Transients}. We model this data with an established stellar synthesis code with a star formation history appropriate for the lensed disk galaxy, a.k.a. the Dragon Arc, using {\sc SPISEA} (Stellar Population Interface for Stellar Evolution and Atmospheres \citep{Hosek_2020}) to trace the stellar evolution based on a star formation history suited to the spiral galaxy that is lensed into the Dragon Arc, given its redshift and optical-IR photometry. This star formation history is estimated through SED fitting with {\sc Bagpipes} \citep{Carnall_2018}, with a Kroupa IMF \citep{Kroupa}. A more detailed discussion of the methodology, including the effect of choice of different model parameters, can be found in Li et. al. (2024, in prep).
   
  The model-generated stellar population generated at any given time step is then input into our microlensing calculations with a predefined set of probability distributions (PDF) calculated numerically as a function of both the stellar surface density in the lens and the level of macro magnification of the lensing host of the microlenses, as outlined in \citep{Palencia_23}. It is notable that even the modest, commonly observed percent levels of projected stellar mass compared to the total column of dark matter near the Einstein radius of a cluster, $\kappa_\star/\kappa_{ER}\simeq 0.01$, can generate trails of multiply microlensed star images that appear and disappear on a wide range of timescales in the optically thick microlensing region close to a cluster critical curve \citep{Diego_2018}. 
  
  \begin{figure}[htb!]
 \gridline{\fig{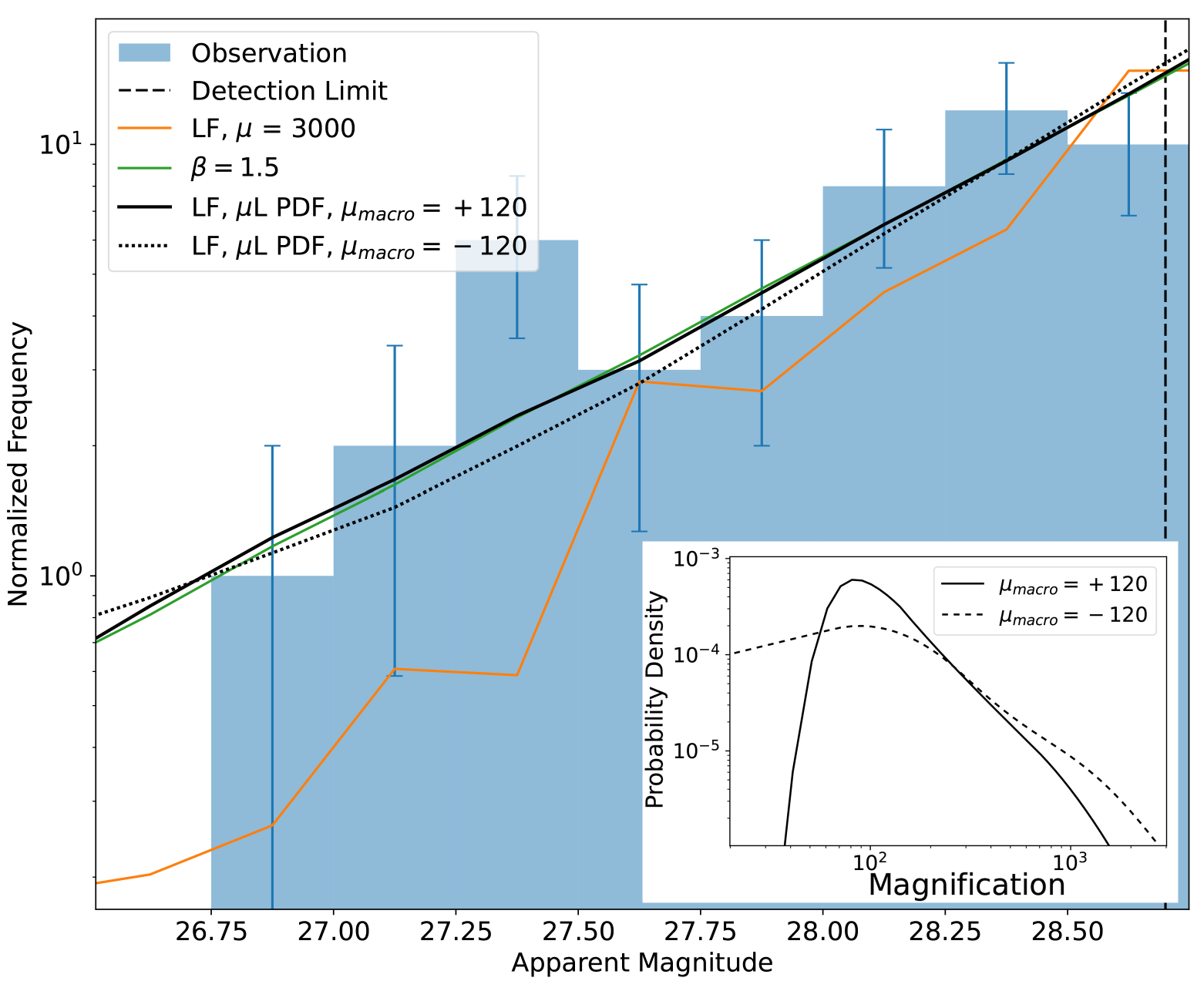}   {0.45\textwidth}{}}
    
    \caption{Microlensed events detected by JWST, with magnitudes measured in the F200W band of JWST, compared with the predictions for the stellar population in an evolved spiral galaxy. The black curve includes microlensing by the stars comprising the diffuse cluster light. This accounts well for the brightness distribution of microlensed stars above the flux limit indicated by the vertical black line and
    the model predicts these are mainly red giant AGB stars. The orange curve shows the unlensed stellar distribution simply shifted by a mean magnification, $\mu=3000$ or $7.5$mags, approximately matching the data. The green power-law fit indicates the counts slope is $d\log N(m)/dm\simeq 0.7$, well above the lensing invariant slope ($\gamma=0.4$), i.e. a positive magnification bias and thus a much larger number of microlensing events is potentially accessible in modestly deeper images of the Dragon Arc. These calculations account for the parity dependence of the magnification distribution shown in the inset, in particular the longer tail to higher magnification that allows lower luminosity stars to be detected. The curves in the inset are for the choice of a mean "macro magnification" of $\mu=\pm 120$ corresponding to the mean magnification that "underlies" the extra microlensing predicted at their observed locations relative to the cluster critical curve - as seen in Figure~\ref{fig: DragonArc_Transients}.} 
    \label{fig: JWST_Histogram_AGBLF}
\end{figure}

 The ICL starlight at the location of the Dragon Arc translates to a surface mass density of about $18\, M_{\odot}/pc^2$, for the Kroupa IMF, or about 1\% of the total column of dark matter near the Dragon arc. With this we can generate statistical distributions of microlensed stars for comparison with the observed JWST transients in the F200W band, $m_{AB}<28.6$. We can now predict the number counts, $N_L(<m)$, of microlensed events as a function of lens magnification, relative to the unlensed counts at fixed apparent magnitude, $N_o(<m)$. Predictions for the microlensing rate must include this magnification bias \citep{BTP_1995}:
  \begin{equation}
    N_L(<m)/N_o(<m)=\mu^{2.5\gamma(m)-1.}
  \end{equation}
  
   showing the competition between the reduced source plane area, $\mu^{-1}$ and enhanced magnified depth at the apparent magnitude, $m+2.5\log\mu$, leading to a net enhancement in numbers, or positive magnification bias, when the slope of the counts is steeper than the break-even value,  $\gamma(m)=d\log N(<m)/dm > 0.4$. This magnification bias is sizeable for the steep bright end of the observed counts, $\gamma \simeq 0.7$ as shown in Figure~\ref{fig: JWST_Histogram_AGBLF}, where we compare our predictions with the Dragon arc.
   
 We see that the steep counts are well matched by our normalised predictions in Figure~\ref{fig: JWST_Histogram_AGBLF}, where the solid black curve in Figure~\ref{fig: DragonArc_Transients} represents the combined negative and positive parity images, for which we also show there is a small asymmetry, which is small in relation to the count slope but as we show later does significantly differ across the cluster critical curve, shown in Figure~\ref{fig: JWST_Histogram_AGBLF}. For these predicted detections the mean magnification of the microlens and background cluster magnification is close to $\mu \simeq 10^3$ and corresponds to luminous red giants in the Dragon Arc, with absolute magnitudes of $-8<M_{AB}<-6$, in the F200W band of JWST.

\section{Smooth Dark Matter plus stellar microlenses.}

 The simplest combination of stellar microlensing added to a smooth distribution of dark matter has been explored in detail since the discovery of ``Icarus", the first microlensed star at cosmological distances, a blue supergiant in the spiral arm of a lensed galaxy at $z=1.51$ \citep{Kelly_2016} near the critical curve of the massive cluster MACS J1149($z=0.54$). A relatively modest projected density of stellar mass is now appreciated to be sufficient for generating the microlensing of Icarus and other transient stars near the critical curve of other lensing clusters with a projected mass density typically observed to be  $\kappa_\star/\kappa_{ER}\simeq 0.01$, at the Einstein radius (ER) of the cluster, where the scale of microlensing is effectively magnified by the macro-lensing of the cluster to become the main source of magnification: 
 
 Numerical calculations with this one percent level of microlensing show the critical curves of the microlensing stars are effectively magnified by the tangential magnification factor $\mu_t$, becoming a dense web at high optical depth, $\tau>1$. In practice this transition occurs for $\mu_t\kappa_* > 1$, with a typical stellar density of microlenses $\kappa_*=0.01$, corresponding to $\mu > 100$, where $\mu=\mu_r\mu_t$. Near the Einstein radius $\mu_r\simeq 1$ for a projected DM profile of $d\log \kappa/d\log r \simeq -1$, for the isothermal case and a somewhat shallower for NFW within the scale radius, $r_s$, depending on the concentration, $c_{NFW}$, so $r_s=R_{vir}/c_{NFW}$. We can see that most of the observed lensed events do indeed fall within this range $|\mu|>100$ that we outline in green in Figure~\ref{fig: DragonArc_Transients}, spanning a band of $\simeq 4$kpc about the predicted critical curve of A370.

 \begin{figure*}[htb!]
    \gridline{\fig{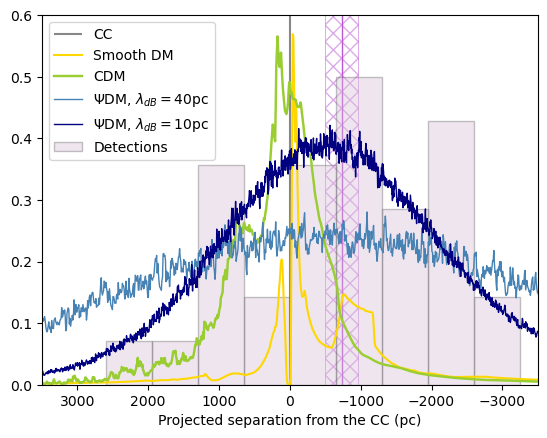}   {0.45\textwidth}{} \fig{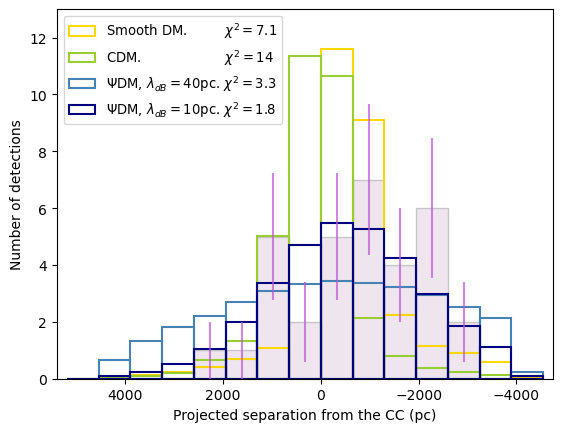}{0.45\textwidth}{}}
    \caption{The offset distribution of JWST transients (33 detections) along all four intersections of the Dragon Arc traversed by tangential critical curve of A370, labeled A,B,C,D in Figure~\ref{fig: DragonArc_Transients}. The left panel shows the form of the predicted distributions for comparison with the asymmetric distribution of the data which favours negative offsets interior to the cluster critical curve, with a mean of $-0.7\pm0.2$kpc shown by the hatched vertical band. The broad spread with negative skewness of the observed histogram resembles best $\psi$DM, with a de Broglie scale of 10pc favoured with $\chi_r^2=1.8$, as shown on the right, compared to $\psi$DM with a larger de Brogle scale of 40pc.The data are in significant tension with the centrally peaked smooth DM model shown in yellow, and with the CDM based sub-halo model, in green, as detections are skewed positive, unlike the data.}
    \label{fig: Histogram_comparison}
    \end{figure*}
 
 The ideal critical curve for a smooth cluster lensing profile is replaced by a dense corrugated band of critical curves centered on the cluster critical radius. Then the peak magnification of background stars occurs when any microlensing caustic sweeps past the observer, mainly from the transverse peculiar velocity of the lensing cluster, highly magnifying small background sources. The peak magnification of a star of finite radius is set by the density of microlensing network, i.e. by the optical depth of microlensing, $\tau$, which is proportional to the product of $\mu_t\Sigma_{\star}$. This generates peak levels of microlensing as the microlensing caustic network crosses the lensed stars reaching $\mu_{max} \simeq 10^{3-4}$ for $\kappa_{\star} \simeq 0.01 \kappa_{ER}$ lasting hours to days depending on the star radius and relative transverse velocity, $v_t$, simply $t_{\textrm{peak}}=2R_{\star}/v_{t}$, and this will be repeated for a given star on average every few years at high $\tau$ \citep{Venumadhav_2017, Diego_2018, Oguri_2018}. 
 
 The width of this dense micro lensing band is set by the separation where the optical depth exceeds unity, $\tau>1$. Ray tracing at high resolution has uncovered an important difference in the distribution of microlensing magnification due to the switch in the sign of the local orthogonal stretch factors, from $1-\kappa+\gamma$ to $1-\kappa-\gamma$ at this radius,  which causes the radially directed "figure of 8" shaped critical curve of a point mass on the outside to being tangentially aligned with the critical curve on the inside and more magnified by the boost in magnification parallel to the cluster curve \citep{Venumadhav_2017, Diego_2018}. Recent high-resolution simulations described above \citep{Palencia_23} have now provided more detailed predictions for the distribution of magnifications as a function of the underlying macro-magnification and the projected surface mass density of microlensing stars. This parity dependence is shown in the inset in Figure~2 and allow a reliable comparison with the observations for the high optical depth regime $\tau>1$,  where numerical predictions are essential and have demonstrated an important distinction between the image parity regimes inside and outside the critical curve, where a longer tail to high negative magnifications is predicted for microlensing with $\tau>1$ which leads to a higher detection rate on the inside of the cluster critical curve as we emphasise here and shown in the inset of Figure~\ref{fig: JWST_Histogram_AGBLF}. Also as understood in other numerical work, for both parities the probability $P(\mu)$ peaks at $\mu_{peak} = 10^{3-4}$, being smaller at higher $\tau$, rather than $\mu\simeq 10^{6}$ for well-separated microlensing stars stars when $\tau<1$.

\section{Adding dark CDM sub-halos}

\begin{figure*}[ht!]
    \centering
    \includegraphics[width = \textwidth]
    {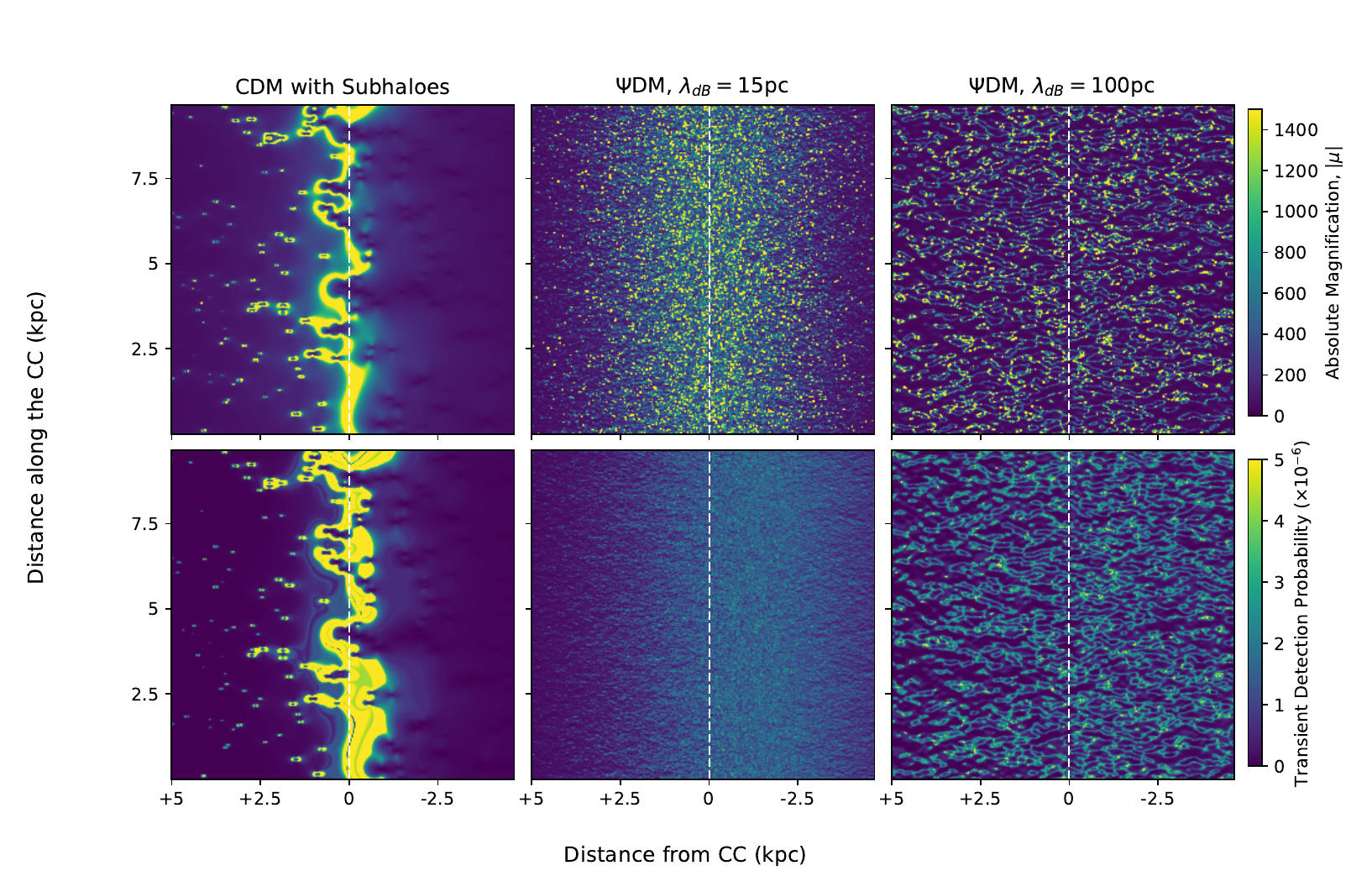}
    
    \caption{Lensing Substructure Predictions: The top row shows the dark matter magnification including substructure for CDM (top left) and the de Broglie waves for $\psi$DM, for two choices of de Broglie wavelength, 15pc and 100pc (top center and top right respectively). The lower row is the microlensing detection rate based on the magnification map from the top row with the microlensing magnification included and convolved with the lensed star luminosity function. For CDM the sub halo masses range is $10^{6-8}M_\odot$, with $C_{NFW}=30$ and critical curves can be seen around sub-halos mainly on the outside of the cluster critical curve. The microlensing detections are seen to trace these critical curves, enhanced by the magnification bias (bottom left). We predict reduced negative magnification locally by the sub-halos interior to the cluster critical curve and hence a deficit of microlensing. For $\psi$DM a broad network of critical curves is formed symmetrically about the Einstein radius of the cluster (dashed vertical line) which for $\lambda_{dB}=10$pc ranges over a $\simeq 4$kpc band (top center), like the data. A much larger spread is predicted for $\lambda_{\psi}=100$pc de Broglie (top right), spread over $\simeq 40$kpc, much broader than the observations. The corrresponding microlensing detection rate is seen to be shifted to the inside of the cluster critical line for $\psi$DM by about $-0.6$Kpc for $\lambda_{\psi}=10$pc(center bottom), like the data, caused by the longer tail to higher magnification for negative parity images. The opposite shift is predicted for CDM microlensing detections as they are skewed into the outside the Einstein radius following the extra sub-halo critical curves formed there (bottom left).} 
    \label{fig: MagMap_TrProb}
\end{figure*}

   The breadth of the microlensing region can be increased by a population of dark halos of sufficient mass, as explored by \citet{williams2023flashlights} in the context of the Flashlights program, showing there is a distinct preference for new critical curves to form around halos that lie outside of the critical curve \citep[see also][]{Abe_2024}. We follow this prescription including a power-law distribution of sub-halo masses with $m_{halo}^{-0.9}$  in the relevant mass range $10^{6-8}M_\odot$ \citep{Dai_2020}, for which we adopt concentrated NFW profiles for these dark halos, with $C_{NFW}=30$ appropriate for their relatively low masses \citep{Navarro_1997}. We see in Figure~\ref{fig: Histogram_comparison} that such a population of dark halos CDM can be chosen such that to broaden the critical region as desired by several kpc, but with a clear preference for the new critical curves to appear outside the critical radius of the radius of the cluster where the excess sub-halo mass can exceed the critical density for lensing and thereby generate local Einstein rings around the sub-halos located there. The opposite behaviour is predicted on the inside of the critical curve, as the critical density is already exceeded and instead locally reduced magnification is predicted for each sub-halo and thus local critical curves are not generated, as can be seen in Figure~\ref{fig: MagMap_TrProb}. Also shown in Figure~\ref{fig: MagMap_TrProb}, is the addition of microlensing described by the magnification dependent probability densities described in \S1 for the high optical depth $\tau$ regime, with the detection limit of $m_{AB}=28.3$ set by the JWST observations of the Dragon Arc. This is shown in the lower left panel of Figure~\ref{fig: MagMap_TrProb}), where it can be seen that the critical curves remain visible above a smoother background of transients thus skewing detectable microlensing to the outside of the cluster critical curve, unlike the data where the opposite tendency is apparent.

\section {Wave Dark Matter}

This lack of observed positive skewness above for CDM-like dark halos motivates us to consider Wave/fuzzy Dark Matter, $\psi$DM as a simple, desirable proposal for the dark matter 
\citep{Schive_2014,Hui_2017, Niemeyer_2020, Hui_2021} given the now stringent absence of WIMPs and the natural solution that $\psi$DM provides for the small-scale problems of CDM, with its inherent Jeans Scale \citep{Schive_2014,Niemeyer_2020}. Simulations in this context have revealed a surprisingly rich wave structure due to interference, including a central soliton that is widest at low momentum, providing a natural explanation for the large cores of common dwarf Spheroidal galaxies \citep{Schive_2014,Pozo_2024}. Unique lensing effects are also to be expected, due to the full density modulation of the $\psi$DM in cluster halos that leads to a wide corrugated band of critical curves all along the Einstein radius of a lensing galaxy or cluster, at a level that depends on the projection of independent de Broglie scale fluctuations along the line of sight, resulting in a Gaussian random field (GRF) of perturbations,  by the central limit theorem. The surface density at a projected radius $R$, $\Sigma(R)$, is perturbed by, $\Delta\Sigma(R_h) / \Sigma(R_h) \simeq (\lambda_{dB}/R_h)^{0.5}$, where $R_h$ is the effective halo size at the Einstein radius \citep{Venumadhav_2017,Kawai_2022}. This corresponds to a 1\% dispersion relative to the mean surface mass density at Einstein radius of massive cluster where, $R\simeq 100$kpc and a de Broglie scale of $\lambda\simeq 10$pc is predicted (see below). Interestingly, this says the perturbation amplitude increases towards the cluster center as there are fewer independent cells of $\lambda_{dB}$ relative to the projected radius, R and hence larger density perturbations of 2-3\% are expected near the radial critical curve with strong lensing effects. 

For clusters, the de Broglie Wavelength should be relatively small, scaling with the widely adopted core-halo scaling relation \citet{Schive_2014b},
\begin{equation}
\lambda_{dB}=15\,\left( \frac{10^{-22}\, {\rm {eV}}}{m_{\psi}} \right) \left( \frac{10^{15}\, M_{\odot}}{M_{\rm {Halo}}}\right)^{1/3}\, \, {\rm pc}\, 
\end{equation}
where $m_{\psi}$ is the boson mass of $\psi$DM. For masses $m_{\psi} \approx 10^{-22}$\,eV and a $10^{15}\,M_{\odot}$ cluster, and corresponds to 3\,mas in the lens plane. The Uncertainly Principle 
proves a simpler scaling when using velocity dispersion,  which for dSph galaxies is typically $\simeq 10$km/s 
and for massive clusters $\simeq 1000$km/s, so $\lambda_{clus}/\lambda_{dSps}=({\sigma_{clus}/\sigma_{dSph}})^{-1}=0.01$, i.e. we simply expect, $\lambda_{clus}\simeq 10$pc. We can also predict the width of the corrugated band, $\Delta_{\psi}$,  as a function of $\lambda_{dB}$ and the observed Einstein radius of the cluster, $\theta_{E}$, and with the GRF condition above we have  $\Delta\Sigma_E \sim \Sigma_E \sqrt{\lambda_{db}/R_h}$, and we can adopt $R_h\sim$ 100pc at the projected radius of the Dragon Arc and surface mass density at the Einstein radius, $\Sigma_E/ \Sigma_{crit}=0.7$ for the NFW profile,

\begin{equation}
\Delta\theta_{\psi} \simeq \theta_E\times 0.0022 \sqrt{\lambda_{db}/1pc}
\end{equation}

For $\lambda_{db}=15$pc, and an Einstein radius of approximately, $30\arcsec $ for A370 at the redshift of the Dragon Arc this predicts a full width of $\Delta\theta_{\psi} = 0.52\arcsec $ or $\Delta_{\psi}=3.8$kpc,  which we see in Figure~3 is indeed close to the observed width. In addition to this band of critical curves we add the micro-lensing from starlight in the ICL visible at the location of the Dragon arc, described above with $18M_\odot/pc^2$ \citep{Meena_2023_offcaustic} or about 1\% of the dominant dark matter column at the Dragon Arc location. The upper row of Figure~\ref{fig: MagMap_TrProb} shows the magnification pattern for two choices of boson mass, including the prior predicted 10pc scale based on local dSph dwarf galaxy cores, and also a larger 100pc scale where more detail of the corrugated pattern is visible. As can be seen in 
Figure~\ref{fig: MagMap_TrProb} the width of the magnified band of critical curves, $\Delta_{psi}$, is about $4.5$kpc for $\lambda_{\psi}=10$pc, similar to the data, but much larger for the 100pc scale, and is predicted to scale as, $\Delta_{\psi} \propto \sqrt{\lambda_{dB}}$.

The effect of microlensing is to bias the detections to smaller radius, inside the critical curve of the cluster, at a level of about $-0.53$kpc for $\lambda_{dB}=10$pc, as seen in the lower central panel of Figure~\ref{fig: MagMap_TrProb}, and similar to the observed offset of $-0.7\pm 0.2$kpc for the data, indicated in Figure~\ref{fig: Histogram_comparison}. This asymmetry in the microlensing band relative the symmetric band of the critical curves follows from the higher magnification for negative parity images and the positive magnification bias for red giant stars (see Figure~\ref{fig: JWST_Histogram_AGBLF}). It is important to appreciate that
the agreement we find here is for the prior predicted de Broglie scale of $\simeq 15$pc is based on independent local dwarf galaxy core fitting to the soliton profile finding a boson mass of $\simeq 10^{-22}$eV is preferred for the dominant dark matter in the $\psi$DM context \citep{Schive_2014,Chen_2017,Pozo_2024}.

\section{Discussion and Conclusions}

We have shown the abundant microlensing in the Dragon Arc detected by JWST \& HST follows closely the tangential critical curve of A370. We emphasise that the path of the critical curve is accurately defined by the reflection symmetry of many internal features now recognisable within the Dragon Arc, as shown in Figure~\ref{fig: DragonArc_Transients}. Furthermore, this model-independent path is almost indistinguishable to our free-form adaptive grid-based lens model WSLAP+ \citep{Diego2007,Sendra,Diego2024_3M} built from over 90 multiply-lensed galaxy images around A370 \citep{A370_model} also the LTM (light traces mass) method \citep{ZB} can be seen to provide a reasonably good comparison with the observed distribution \citep{Fudamoto_2024}. Hence, we have accurately pinned down the path of the critical curve allowing a precise comparison with the microlensing events and the tight correspondence means higher magnification is required for detection. This we estimate to have a mean level of $\mu\simeq 3000$ by convolving our stellar synthesis code with the microlensing probability distribution. We therefore disfavour interpreting these transients as very luminous giants with modest magnification, for which detections would be more uniformally spread along the Dragon Arc. Instead, most of the JWST detections we conclude correspond to AGB stars with absolute magnitudes in the range, $-6 <M_{F200W} <-8$, which reproduces well the observed counts that all lie faintward of F200W$>$26.6, with rapidly rising numbers towards the detection limit of F200W$=$28.6. The surprisingly large numbers of microlenses then can be understood as these abundant AGB stars are accessible due to the low redshift of the Dragon Arc($z=0.735$) and the deep JWST data. It is also clear that a huge reservoir of RGB stars sits just below F200W$>$28.5 that modestly deeper JWST data can access. 

 We have noticed an apparent asymmetry in the locations of the microlensed stars of $-0.7\pm 0.2$kpc, favouring detection along the inner edge of the critical curve with a 2:1 ratio. This is very interesting as images of negative parity are in fact expected to reach higher magnification than on the positive parity side, thereby reach lower luminosity and hence more numerous stars in the Dragon Arc. We have calculated the level of skewness by combining a stellar synthesis model with our microlensing simulations that extends to a high optical depth regime of relevance here. We predict a small negative skewness of $-0.05$kpc for the combination of smooth underlying dark matter plus stellar microlensing at the observed level (given by the ICL near the Dragon Arc), which is modest compared to the skewness observed, $-0.7\pm 0.2$kpc. Furthermore, microlensing should be sharply peaked in a narrow band of only $1.4$kpc, compared to the $\simeq 4$kpc wide band observed. This motivates adding invisible substucture on a larger milliarcsecond scale, such as CDM subhalos or the de Broglie scale of $\psi$DM, to generate a wider band of critical curves that microlensing detections can follow.
 
 We have shown that CDM substructure with a mass rage of $10^{6-8}M_\odot$ can widen the band of microlenses significantly with subhalos in the range $10^{6-8}M_\odot$ but detections are predicted mainly the outside of the critical curve as sub-halos in this region can exceed the critical density for lensing and thereby generate a local Einstein ring, whereas inside the critical density the critical density is already exceeded and we have found the opposite behaviour occurs with a dip in magnification generated locally by each sub-halo. Clearly, the negative skew of the Dragon Arc disfavours such CDM-like sub-halos, and this prediction will be more stringently tested with additional cadenced JWST imaging of the Dragon Arc, now underway. Our constraint on CDM complements the upper limit on sub-halos in a recent search for the source of flux anomalies for quad-images of AGN lensed by massive galaxies in new JWST/MIRI imaging, where the absence of locally perturbing CDM-like sub-halos constrains any such population to lie below $< 10^{7.3}M_\odot$,
 \citep{Nierenberg_2024},

  \begin{figure}[htb!]
 \gridline{\fig{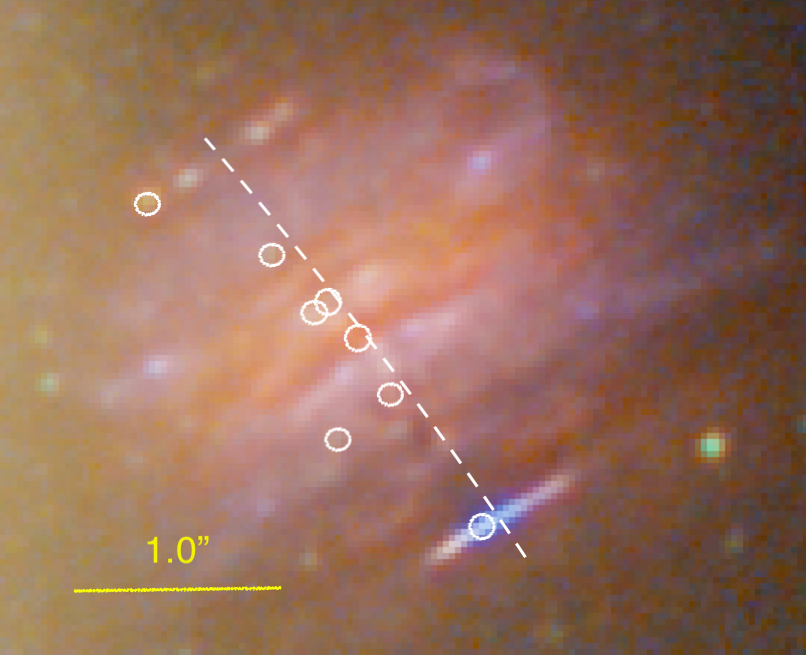}   {0.45\textwidth}{}}
    
    \caption{Microlensed stars detected by JWST in this multiply-lensed galaxy ($z=0.94$) behind the massive cluster MACS J0416, observed by the PEARLS team \citep{Yan2023webbs,Windhorst_2023}, with an additional transient star detected by HST ``Warhol" \citep{Chen_2019} - bottom circle. The mirror symmetry is very clear and reveals that the 8 microlensed stars (white circles) favour the left side of the lensed galaxy, following the ``inside" of the tangential critical curve, (dashed white line) with offsets on the sub-arcsecond scale like the Dragon Arc, providing independent confirmation.} 
    \label{fig: Warhol}
\end{figure}

 We have emphasised that for $\psi$DM the relatively wide band and negative skewness of the distribution of microlenses seen in the Dragon Arc is readily accounted for by the equal levels of positive and negative density interference inherent to $\psi$DM. This symmetry is unique to $\psi$DM and generates critical curves that follow locally negative fluctuations interior to the Einstein radius of the cluster \citep{Amruth_2023}, whereas outside this radius the opposite occurs with critical curves locally following positive fluctuations. Generally, for any clumpy dark matter density fluctuations are only positive, including CDM, for which local critical curves are not formed for subhalos inside the Einstein radius of the cluster. We find quantitative agreement with the Dragon Arc for de Broglie scale of about 10pc, corresponding to $m_\psi \simeq 10^{-22}$eV. 
 
 Remarkably this boson mass scale is predicted a priori from the size of local dwarf Spheroidal galaxy cores that match the predicted central soliton ground state of $\psi$DM the radius of which is also given by the de Broglie scale \citep{Schive_2014,Chen_2017,Pozo_2024}. This lensing-based result favouring $\psi$DM, reinforces recent analysis of the milli-arcsecond positions ``anomalies" of high-resolution radio lensing around a massive lensing galaxy H2018+1906 \citep{Hartley}, for which the de Broglie scale of $\simeq 100$pc is favoured and this implies a boson mass of $m_\psi=10^{-22}$eV for the dominant dark matter when converting by the momentum scale \citep{Amruth_2023}. Note that this momentum dependence of the de Broglie scale is important allowing for consistency checks on this scenario, whereas for CDM any substructure should be largely independent of scale but with some radially dependent effect from tidal forces.

In detail, a precise boson mass would require allowance for some variation in the momentum of the dark matter expected within A370, that may affect the de Broglie scale given the unfinished virialization state implied by the observed bimodality of A370, with hydrodynamical modeling indicating an ongoing major merger viewed after two core passages \citep{Umetsu}. for $\psi$DM some allowance may also be made for the presence of massive cluster members as the de Broglie scale is larger, scaling as $M_{halo}^{-1/3}$ \cite{Schive_2014b}, depending on the column density its DM relative to that of the cluster and cause the overall width of the microlensing band to expand locally, scaling as $\sqrt{\lambda_{dB}}$. Also, some level of smoothing of the caustic structure from the presence of the smooth ICL gas may be expected at the 10\% level. We can also improve upon our comparison by taking more account of variations in red giant star density expected over the surface of the Dragon Arc too and this can be examined with our stellar population synthesis code fitted spatially to the Dragon Arc in 2D. Regions of younger star formation will favour the presence of more luminous young giants as although most of the AGB contribution is from the post-RGB phase, this phase can be relatively recent for relatively high-mass red giants because of their short main-sequence lifetimes.
 
 Deeper imaging of the dragon arc with a monthly cadence can uncover an order of magnitude more microlensing events per year, and with deeper imaging to reach into the RGB population, as shown in Figure~\ref{fig: DragonArc_Transients}. Such data can reveal the fine structure of any network of critical curves, in particular any relatively large CDM-like subhalos with their individual Einstein rings, as can be seen in Figure~\ref{fig: MagMap_TrProb}. Radially elongated corrugations and island critical curves of $\psi$DM may then be visible directly, with sufficient numbers of microlenses, particularly for a larger de Broglie wavelength. If with more data a featureless band is found this would indicate microlensing by uniform, compact objects such as PBH or other compact microlensing objects, especially if a prominent central band along the critical curve is seen, as predicted in Figure~\ref{fig: Histogram_comparison}. The light curves can help as high frequency, small flux modulations on a daily timescale can distinguish PBH from the more smoothly varying by a lower level of stellar microlensing modulated by CDM or $\psi$DM substructure. The colour of the microlensing stars can also help distinguish positive from negative parity regions to identify the precise location of the cluster critical curve as the more magnified events along the inside the critical curve reach further into the RGB where the lower luminosity stars are somewhat bluer. It should also be possible to see the corrugated pattern of critical curves for low redshift images lensed by massive galaxies, as the lower momentum means predicts a resolvable de Broglie scale of 100pc, an order of magnitude larger than for cluster lensing. This predicted momentum dependence for $\psi$DM contrasts with CDM which is not expected to show strong scale dependence as the sub-halo population is largely independent of the host lens.

 Finally, we notice similar behaviour in another low redshift lensed galaxy, $z=0.944$ shown in Figure~\ref{fig: Warhol}, where clear mirror symmetry is visible in JWST images for this fold arc, allowing the path of the critical curve to be identified precisely and independent of lens model uncertainties. Seven microlensed stars have been detected in cadenced PEARLS team JWST images of the massive lensing cluster MACS J0416 \citep{Windhorst_2023,Yan2023webbs} and an additional microlensed star ``Warhol", previously detected by HST \citep{Chen_2019}. All these transients closely follow the inner edge of the critical curve, with offsets on the sub-arcsecond scale like the Dragon Arc and the mirror symmetry here means the source star population is identical on both sides, thereby providing independent confirmation of the preference for negative parity microlensing claimed here. 

\section{acknowledgements}
TJB and PM are supported by the Spanish project grant PID2020-114035GB-100
(MINECO/AEI/FEDER, UE). PM also acknowledges financial support from fellowship PIF22/177, (UPV/EHU). PM thanks the Physics department of Hong Kong University for their hospitality. SKL, TL, and GS acknowledge 
support by the Collaborative Research Fund under Grant No. C6017-20G which is issued by Research Grants Council of Hong Kong S.A.R. RAW acknowledges support from NASA JWST Interdisciplinary Scientist grants NAG5-12460, NNX14AN10G and 80NSSC18K0200 from GSFC. 
JMD and JMP acknowledge the support of project PID2022-138896NB-C51 (MCIU/AEI/MINECO/FEDER, UE) Ministerio de Ciencia, Investigaci\'on y Universidades. 
AZ acknowledges support by Grant No. 2020750 from the United States-Israel Binational Science Foundation (BSF) and Grant No. 2109066 from the United States National Science Foundation (NSF); by the Ministry of Science \& Technology, Israel; and by the Israel Science Foundation Grant No. 864/23. This work was supported by JSPS KAKENHI Grant Numbers JP23K22531, 20H05856, 22K21349, JP22J21440.

\bibliography{refs}{}
\bibliographystyle{aasjournal}

\end{document}